\documentclass[a4paper,11pt]{article}
\usepackage{pos}
\usepackage{xspace}

\newcommand{\gammapy}{\texttt{Gammapy}\xspace}
\newcommand{\mars}{\texttt{MARS}\xspace}

\title{Establishing the MAGIC data legacy: adopting standardised data formats and open-source analysis tools}
\ShortTitle{Establishing the MAGIC data legacy}

\manuallySeparateAuthors
\author*[a]{Cosimo Nigro}
\author{\textit{on behalf the MAGIC Collaboration}}

\affiliation[a]{Institut de Física d’Altes Energies (IFAE), The Barcelona Institute of Science and Technology,\\ Campus UAB, Bellaterra, 08193 Barcelona, Spain}

\emailAdd{cosimo.nigro@ifae.es}

\abstract{The standardisation of gamma-ray astronomical data emerged in recent years as a necessity for the future generation of gamma-ray observatories. Nevertheless, adopting a common format for gamma-ray instruments can already benefit the current generation of gamma-ray instruments. As the end of their operations approaches, it provides a natural solution for the production of their data legacy. Additionally it eases data combination for multi-instrument analyses, thus enhancing the potential for scientific discovery with the wealth of data so far gathered. In this contribution, we present for the first time the effort to adapt the data of the MAGIC telescopes to the standardised format proposed by the \textit{Data Formats for Gamma-ray Astronomy} initiative. We validate the data conversion by analysing the standardised data with the open-source software \gammapy and comparing the results obtained against those produced with the MAGIC proprietary software, \mars. For both sample chosen (Crab Nebula and Mrk421 observation), for all the scientific products extracted (spectra and light curves), we observe good agreement between the results of the two software.}

\FullConference{%
  7th Heidelberg International Symposium on High-Energy Gamma-Ray Astronomy (Gamma2022)\\
  4-8 July 2022\\
  Barcelona, Spain\\}

\begin{document}
\maketitle
\section{Introduction}

The scientific exploitation of very-high-energy (VHE, $E > 100\,{\rm GeV}$) gamma-ray data has been traditionally conducted with proprietary data and analysis software. In recent years, standardised data formats and open-source analysis tools for gamma-ray astronomy \cite{nigro_2021} have been developed in anticipation of the next generation of ground-based gamma-ray telescopes, that will open their observational time and data to the astronomical community \cite{cta_book}. But the current generation of VHE instruments can already profit from this new approach towards data and software. Standardised data provide indeed a natural solution to preserve and make publicly available their observations beyond the end of their operations. If this data legacy were to be made public, its compatibility with open-source analysis tools would be necessary for its profitable usage by the community. Of particular relevance in the implementation of this new open approach in gamma-ray astronomy are: the \textit{Data Formats for Gamma-ray Astronomy} (GADF) forum \cite{gadf}, a community-based initiative defining a standardised format for data from different $\gamma$-ray instruments; and \gammapy \cite{gammapy}, an open-source python package for their analysis. The current generation of VHE instruments has started to produce GADF-compliant data \cite{hess_dl3_dr1} and their potential to be combined in multi-instrument analyses has already been demonstrated \cite{joint_crab, hawc_dl3}.
\par
In this contribution, we present for the first time the effort to convert the MAGIC data in the standardised GADF format. We perform a validation of the point-like analysis by comparing the results obtained analysing the standardised data with \gammapy against those obtained with the MAGIC proprietary software, \mars \cite{mars}.

\section{Data Conversion}

To produce the MAGIC data in the GADF-compliant format we used \mars to reduce the observations down to a data level containing the events (the air showers) with their estimated energy and direction and a score from a classification algorithm ranking their likelihood of being initiated by a gamma ray. Simulated Monte Carlo (MC) data are reduced to the same data level. A proprietary \texttt{C++} library is then used to extract a list of gamma-like events from the observations, and to estimate the instrument response function (IRF) from the MC. Event lists and IRFs constitute the reduced data level, technically referred to as data level 3 (DL3), containing a minimum (detector-independent) level of information necessary to perform a scientific analysis (e.g. estimating a gamma-ray spectrum), and on which the GADF specifications focus. MAGIC's event lists and IRFs are hence stored in \texttt{FITS} \cite{fits} format, compliant with the GADF specs.
\par
We begin the process of validation from the point-like or one-dimensional analysis. In the latter, the position of the source is known and its extension is considered negligible. The signal is estimated from the events within a circular region, referred to as ON, enclosing the source. The background to be subtracted is estimated from one or more regions, referred to as OFF, with the same offset of the ON region, but symmetric with respect to the camera centre. The ON and OFF counts thus extracted are binned in energy (see e.g. Fig.~\ref{fig:counts_comparison}). A likelihood procedure is then applied to fit the observed counts. Expected counts are computed folding the IRF with an analytical parametrisation assumed to describe the source spectrum (see \cite{piron_2001} for a detailed description of the analysis method).

To validate the estimation of spectra and the light curves, the common final scientific products of a point-like gamma-ray analysis, we selected the following data samples:

\begin{itemize}
    \item $50\,{\rm h}$ of Crab Nebula observations gathered between 2011 and 2012. Of these, $30\,{\rm h}$ were observed with the source sitting at $0.4^{\circ}$ offset from the camera centre (single-offset sample), in the zenith range $[5^{\circ}, 50^{\circ}]$. These represent MAGIC standard data-taking conditions and indeed this data constitute the sample used to estimate the MAGIC stereoscopic performance \cite{magic_performance}. In the remaining $20\,{\rm h}$, the source was observed at different camera offsets: $[0.20^{\circ}, 0.35^{\circ}, 0.40^{\circ}, 0.70^{\circ}, 1.00^{\circ}, 1.40^{\circ}]$ (multi-offset sample), in the zenith range $[5^{\circ}, 35^{\circ}]$. These observations were meant to test the response along the whole camera and were used to validate the spatial likelihood analysis method in \cite{skyprism};
    \item $42\,{\rm h}$ of Mrk421 observations from 2013, described in \cite{mrk421_2013}. The source was observed at $0.4^{\circ}$ offset, in the zenith range  $[5^{\circ}, 70^{\circ}]$.
\end{itemize}

The Crab Nebula represent the science case of a steady bright source (optimal to test the spectrum estimation), while Mrk421 that of a bright source with a variable emission in time (optimal to test the computation of a light curve).

\section{Validation}

\begin{figure}
    \centering
    \includegraphics[scale=0.45]{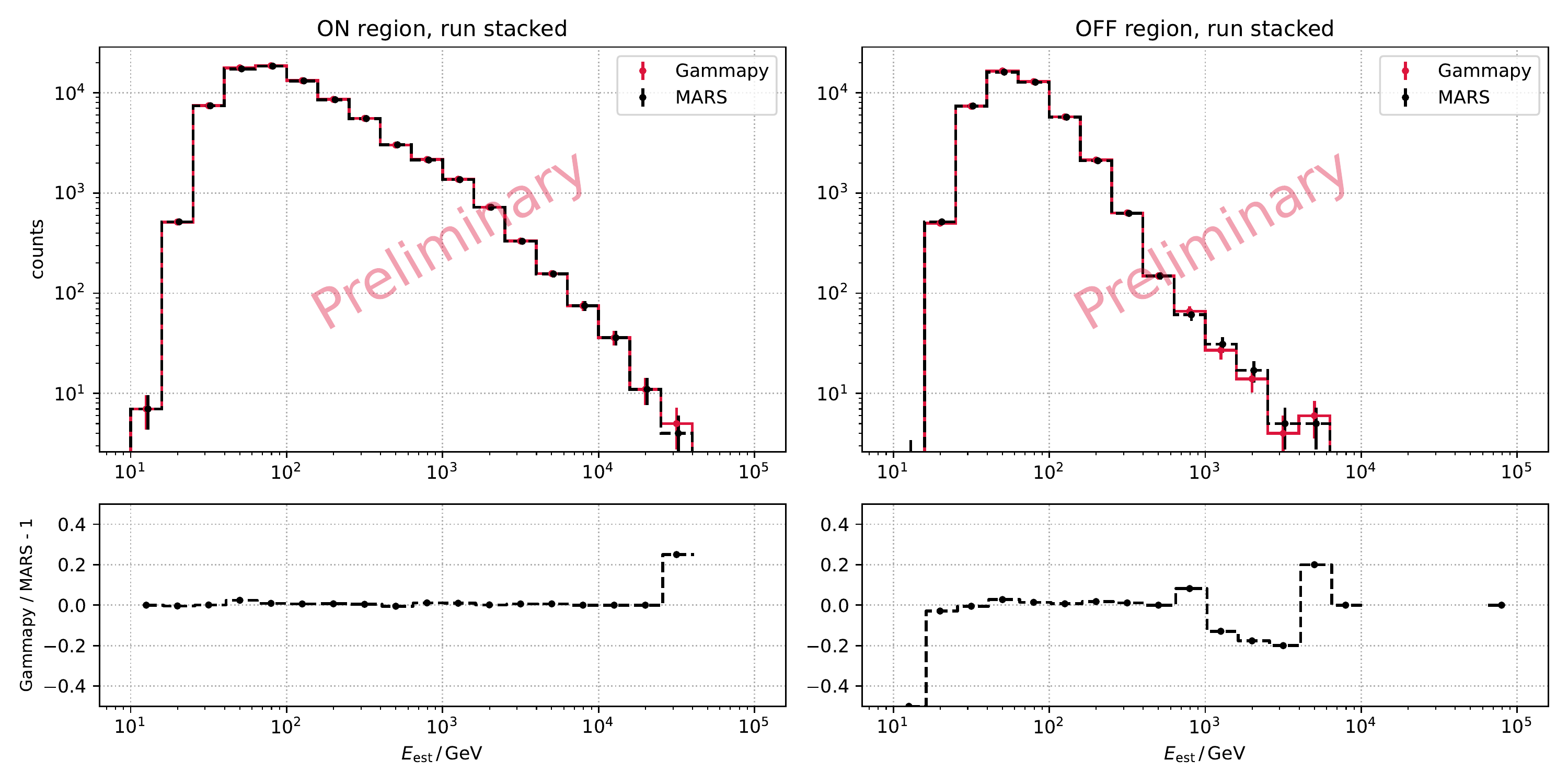}
    \caption{Validation of the signal / background extraction: gamma-ray events in the ON and OFF regions extracted with \mars (black) and \gammapy (red) from the single-offset Crab Nebula sample.}
    \label{fig:counts_comparison}
\end{figure}

The results of the signal / background estimation obtained for the single-offset Crab Nebula sample with \mars and \gammapy are shown in Fig.~\ref{fig:counts_comparison}. We observe an excellent agreement between the ON and OFF counts extracted with the two software, concluding that the information contained in the \mars proprietary data has been correctly translated into the GADF-compliant format. We then estimate the Crab Nebula spectrum using both software and assuming a log-parabolic function to describe the spectrum (Eq.~1 in \cite{magic_performance}). 

\begin{figure}
    \centering
    \includegraphics[scale=0.5]{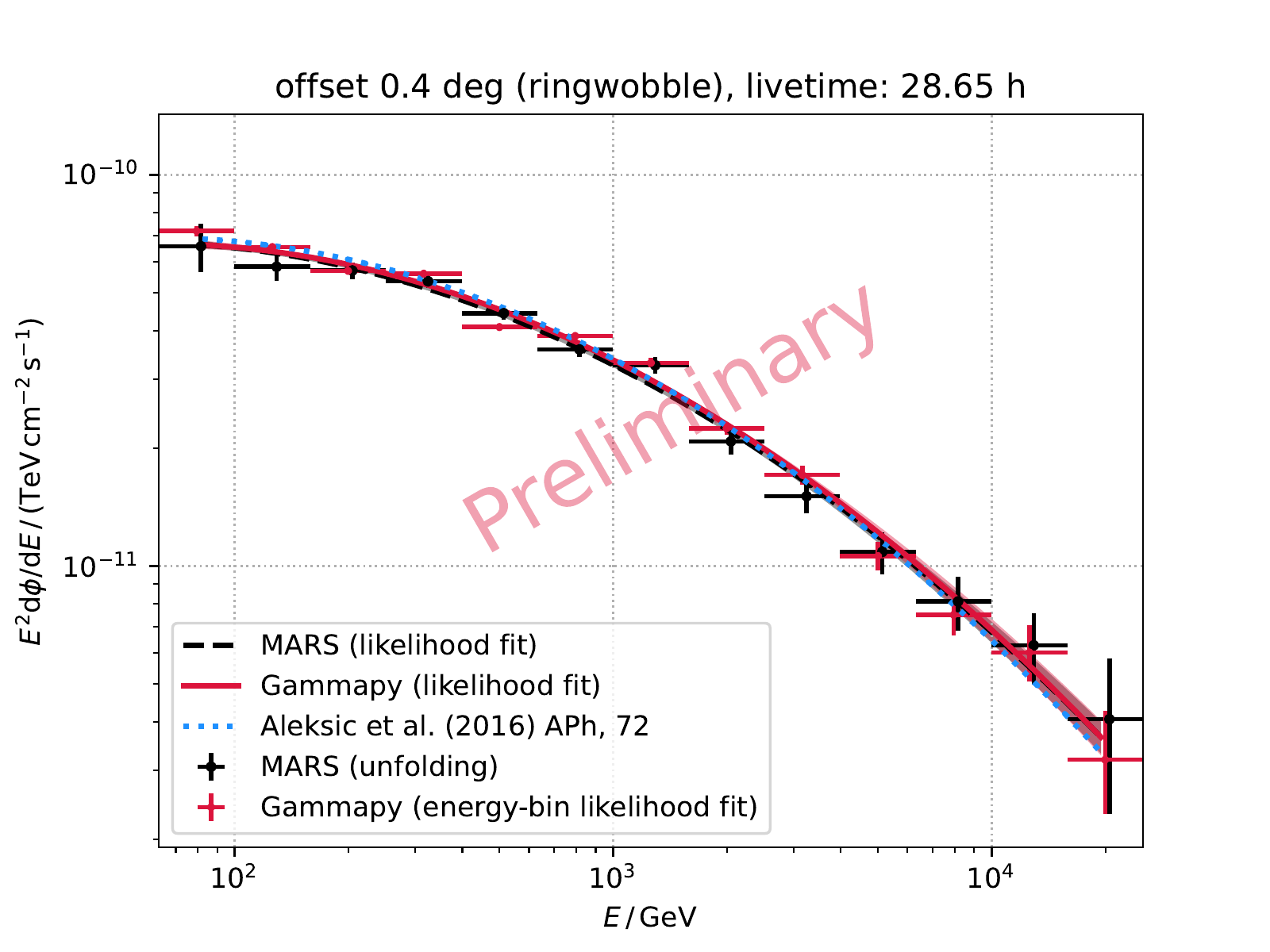}
    \caption{Validation of the spectrum estimation: spectral energy distribution measured from the single-offset Crab Nebula sample using \mars (black) and \gammapy (red). The continuous line and the associated error band represent the result of the forward-folding likelihood fit. Flux points are obtained using the unfolding procedure for \mars, while repeating the forward-folding likelihood fit in a single energy bin for \gammapy.}
    \label{fig:spectra_comparison_single_offset}
\end{figure}

\begin{figure}
    \centering
    \includegraphics[scale=0.42]{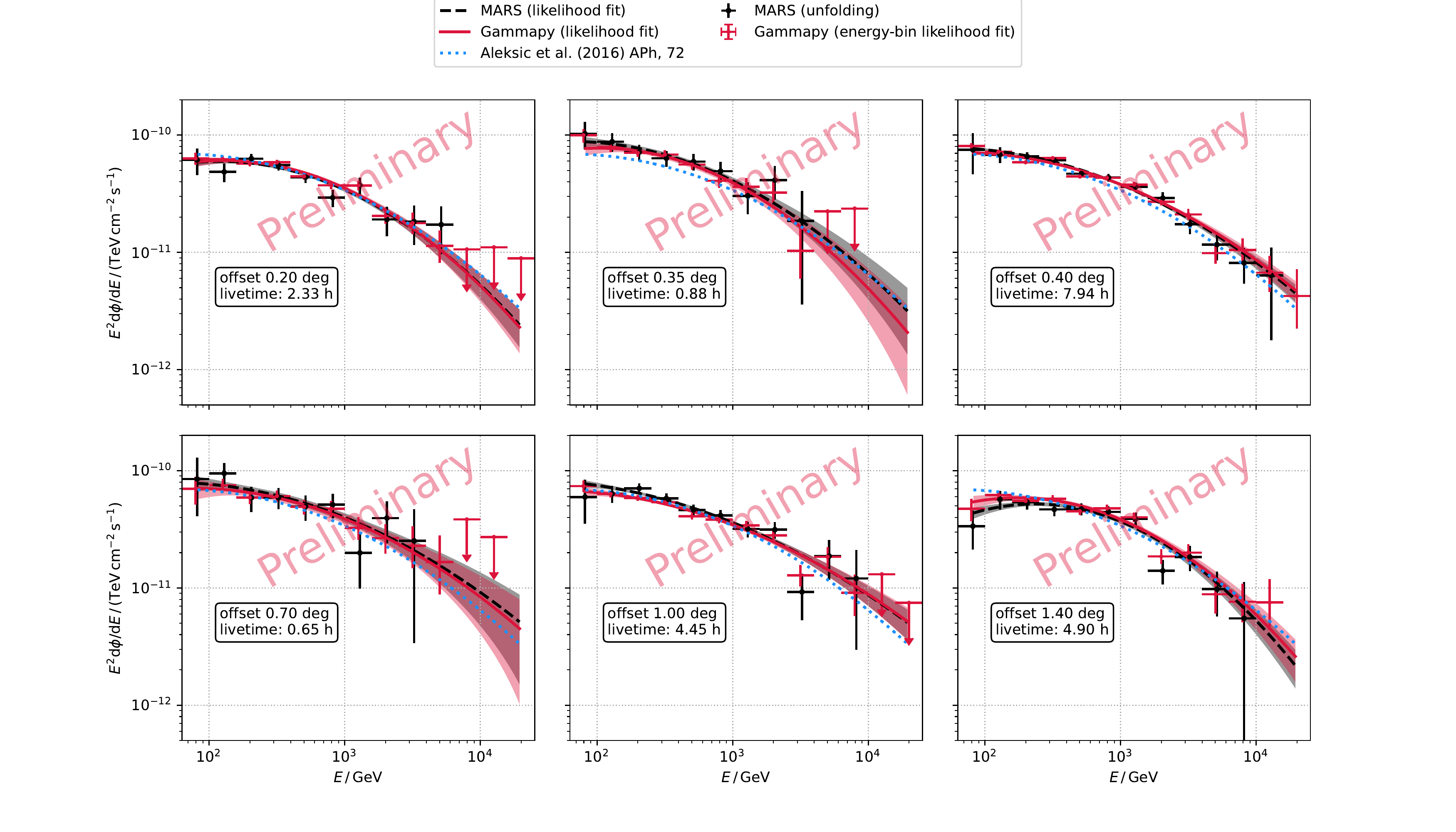}
    \caption{Same as Fig.~\ref{fig:spectra_comparison_single_offset}, but using Crab Nebula observations at different offsets from the camera centre.}
    \label{fig:spectra_comparison_all_offsets}
\end{figure}

Fig.~\ref{fig:spectra_comparison_single_offset} and \ref{fig:spectra_comparison_all_offsets} show the results obtained with the single- and multi-offset samples. In both cases the results of the forward-folding likelihood analysis, described in \cite{piron_2001} and implemented in both software, are in excellent agreement. The flux point computation also returns compatible results despite the two different methods adopted in the two software: unfolding for \mars \cite{magic_unfolding} and a likelihood fit performed with the events in each energy bin for \gammapy. 

\begin{figure}
    \centering
    \includegraphics[scale=0.46]{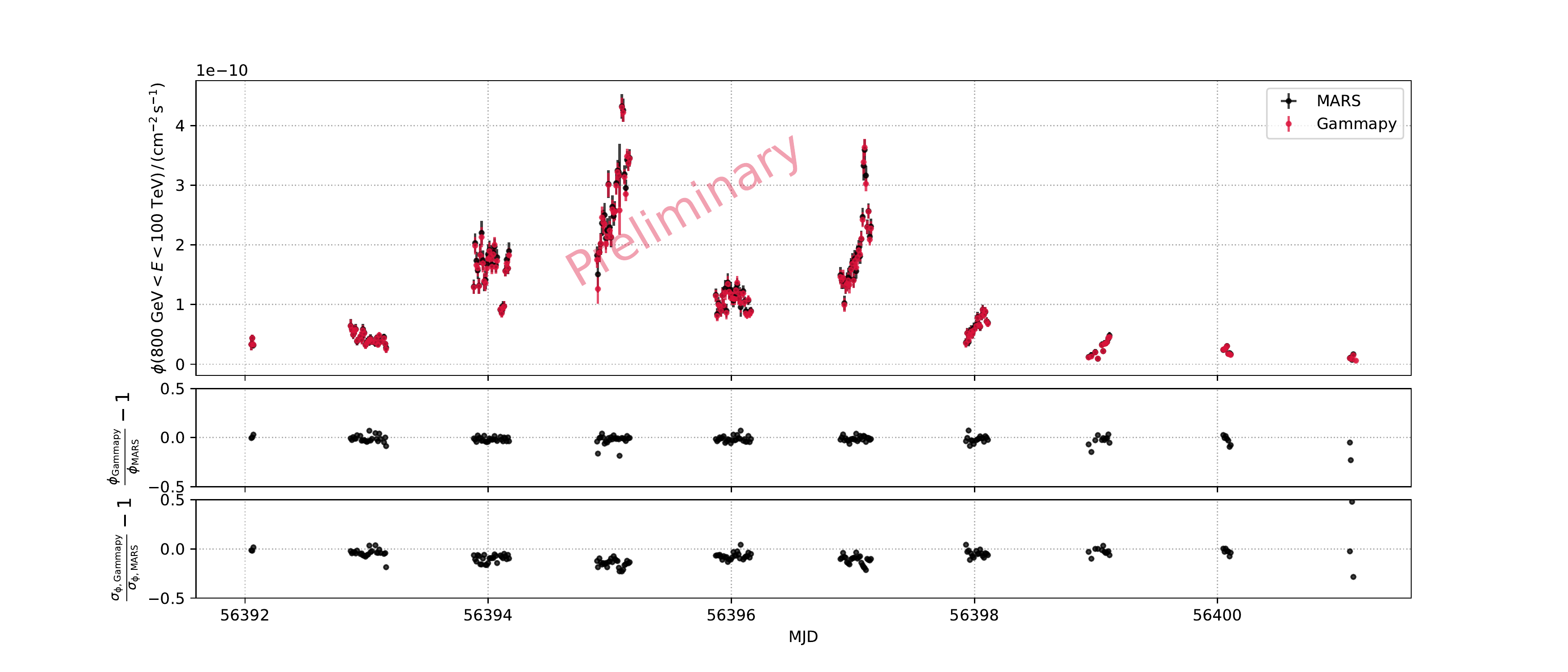}
    \caption{Light curve showing the integrated flux of Mrk421 above $800\,{\rm GeV}$ obtained with \mars (black) and \gammapy (red).}
    \label{fig:light_curve_mrk421}
\end{figure}

We use the Mrk421 April 2013 data to compare the light curve estimation performed by \mars and \gammapy. Fig.~\ref{fig:light_curve_mrk421} illustrates the estimated integral flux above $E_0 = 800\,{\rm GeV}$, obtained with the two software. Consistent results are obtained, despite different methods being used for the flux estimation vs time. \mars estimates the integrated flux from the number of excess (ON - OFF) events above $E_0$, dividing it by the effective area (integrated above $E_0$) and time; \gammapy performs instead a likelihood fit with the events in the specific time bin and reports the integral above $E_0$ of the fitted spectrum. 

\section{Conclusion}
We presented the effort to produce MAGIC data in the standardised GADF format. We validated the point-like analysis by comparing the results obtained with \gammapy against those obtained with the standard \mars analysis chain. For both Crab Nebula and Mrk421 observations we find excellent agreement between the spectrum and light curve estimated with the two frameworks. Major productions of GADF-compliant data ($\sim$ years observational periods) have already being initiated by the MAGIC Collaboration, while at the same time its analysers are starting to adopt GADF-compliant data and \gammapy for their scientific analyses. This represents a milestone in the definition of the instrument data legacy as for the first time its observations can be produced in a standardised format analysable with open-source analysis tools.

\section{Acknowledgements}
C.N. acknowledges support by the Spanish Ministerio de Ciencia e Innovación (MICINN), the European Union – NextGenerationEU and PRTR  through the programme Juan de la Cierva (grant FJC2020-046063-I), by the the MICINN (grant PID2019-107847RB-C41), and from the CERCA program of the Generalitat de Catalunya.


\begin{thebibliography}{99}
\bibitem{nigro_2021}
Nigro, C.; Hassan, T.; Olivera-Nieto, L. 
Evolution of Data Formats in Very-High-Energy Gamma-Ray Astronomy. 
\textit{Universe} \textbf{2021}, 7, 374.

\bibitem{cta_book} 
Cherenkov Telescope Array Consortium. 
\textit{Science with the Cherenkov Telescope Array};
World Scientific Publishing Co. Pte. Ltd.: Singapore, \textbf{2019}.

\bibitem{gadf}
Deil, C. et al.
Open high-level data formats and software for gamma-ray astronomy. 
In Proceedings of the 6th International Symposium on High Energy Gamma-ray Astronomy, Heidelberg, Germany, 11–15 July 2016; American Institute of Physics Conference Series. Volume 1792, p. 070006.

\bibitem{gammapy}
Deil, C. et al.
Gammapy - A prototype for the CTA science tools.
In Proceedings of the 35th International Cosmic ray Conference (ICRC2017), Busan, Korea, 12–20 July 2017; Volume 301, p. 766.

\bibitem{hess_dl3_dr1}
H.E.S.S. Collaboration.
H.E.S.S. first public test data release. 
\textit{arXiv} \textbf{2018}, arXiv:1810.04516.

\bibitem{joint_crab}
Nigro, C. et al.
Towards open and reproducible multi-instrument analysis in gamma-ray astronomy.
\textit{Astronomy \& Astrophysics} \textbf{2019}, Volume 625, id. A10, 8 pp.

\bibitem{hawc_dl3}
Albert, A. et al.
Validation of standardized data formats and tools for ground-level particle-based gamma-ray observatories.
\textit{arXiv} \textbf{2022}, arXiv:2203.05937.

\bibitem{mars}
Zanin, R. et al.
MARS, The MAGIC Analysis and Reconstruction Software. 
In Proceedings of the 33rd International Cosmic ray Conference (ICRC2013), Rio de Janeiro, Brasil, 2–6 July 2013; Volume 33, p. 2937.

\bibitem{fits}
Wells, D.C.; Greisen, E.W.; Harten, R.H.
FITS - A Flexible Image Transport System.
\textit{Astron. Astrophys. Suppl. Ser.} \textbf{1981}, 44, 363.

\bibitem{piron_2001}
Piron, F. et al.
Temporal and spectral gamma-ray properties of Mkn 421 above 250 GeV from CAT observations between 1996 and 2000.
\textit{Astronomy and Astrophysics} \textbf{2001}, v.374, p.895-906.

\bibitem{magic_performance}
Aleksić, J. et al.
The major upgrade of the MAGIC telescopes, Part II: A performance study using observations of the Crab Nebula.
\textit{Astroparticle Physics} \textbf{2016}, Volume 72, p. 76-94.

\bibitem{skyprism}
Vovk, I.; Strzys, M.; Fruck, C.
Spatial likelihood analysis for MAGIC telescope data. From instrument response modelling to spectral extraction.
\textit{Astronomy \& Astrophysics} \textbf{2018}, Volume 619, id.A7, 11 pp.

\bibitem{mrk421_2013}
Acciari, V. A. et al.
Unraveling the Complex Behavior of Mrk 421 with Simultaneous X-Ray and VHE Observations during an Extreme Flaring Activity in 2013 April.
\textit{The Astrophysical Journal Supplement Series} \textbf{2020}, Volume 248, Issue 2, id.29

\bibitem{magic_unfolding}
Albert, J. et al.
Unfolding of differential energy spectra in the MAGIC experiment.
\textit{Nuclear Instruments and Methods in Physics Research Section A} \textbf{2007}, Volume 583, Issue 2-3, p. 494-506.

\end{thebibliography}
\end{document}